\documentclass{emulateapj}
\usepackage{amsmath}
\input{epsf}
\submitted{Accepted for publication in ApJ on August 4, 2015}

\newcommand{\at}[2][]{#1|_{#2}}

\shorttitle{Resonant stirring of dwarf galaxies}

\shortauthors{E. L. {\L}okas et al.}

\begin{document}

\title{The resonant nature of tidal stirring\\ of disky dwarf galaxies orbiting the Milky Way}

\author{Ewa L. {\L}okas\altaffilmark{1}, Marcin Semczuk\altaffilmark{1,2}, Grzegorz Gajda\altaffilmark{1}
and Elena D'Onghia\altaffilmark{3,4}}

\altaffiltext{1}{Nicolaus Copernicus Astronomical Center, Bartycka 18, 00-716 Warsaw, Poland}
\altaffiltext{2}{Warsaw University Observatory, Al. Ujazdowskie 4, 00-478 Warsaw, Poland}
\altaffiltext{3}{Department of Astronomy, University of Wisconsin-Madison, Madison, WI 53706, USA}
\altaffiltext{4}{Alfred P. Sloan Fellow}

\begin{abstract}
Using $N$-body simulations we study the tidal evolution of initially disky dwarf galaxies
orbiting a Milky Way-like host, a process known to lead to the formation of dwarf spheroidal galaxies. We focus on the
effect of the orientation of the dwarf galaxy disk's angular momentum with respect to the orbital one and find
very strong dependence of the evolution on this parameter. We consider four different orientations: the exactly
prograde, the exactly retrograde and two intermediate ones.
Tidal evolution is strongest for the exactly prograde and weakest for the exactly retrograde orbit.
In the prograde case the stellar component forms a strong bar and remains prolate until the end of the
simulation, while its rotation is very quickly replaced by random motions of the stars. In the retrograde case
the dwarf remains oblate, does not form a bar and loses rotation very slowly. In the two cases of intermediate
orientation of the disk, the evolution is between the two extremes, suggesting a monotonic dependence on the
inclination. We interpret the results in terms of the resonance between the angular velocity of the stars in the
dwarf and its orbital motion by comparing the measurements from simulations to semi-analytic predictions.
We conclude that resonant effects are the most important mechanism underlying the tidal evolution of disky dwarf
galaxies.
\end{abstract}


\section{Introduction}

The formation of dwarf spheroidal (dSph) galaxies in the Local Group remains an open question but
one of the most promising scenarios for their origin is via the tidal interaction of their disky
progenitors with more massive hosts like the Milky Way. The scenario,
proposed by Mayer et al. (2001) explains the morphology-density relation observed among the dwarfs of the Local Group
and accounts for the non-sphericity of the dSph objects.

The efficiency of the mechanism and its observational
predictions have been investigated in detail by Klimentowski et al. (2009), Kazantzidis et al. (2011) and
{\L}okas et al. (2011, 2012). These studies explored the dependence of the process on a large number of orbital
and structural parameters of the dwarf. The general picture that emerged from these studies is that a disky dwarf
progenitor, once accreted by a massive host, undergoes strong tidal stirring and mass stripping
if the orbit is tight enough.

Typically,
at the first pericenter passage, the disk transforms into a tidally induced bar (for a detailed description
of the properties of such a bar see {\L}okas et al. 2014a).
The bar becomes thicker and shorter in time leading in the end to the formation of a spheroidal stellar component.
The morphological transformation is accompanied by strong changes in the kinematics as quantified by the amount
of ordered to random motion. The latter starts to dominate at some point and at the end the galaxy is pressure
supported.

Among the parameters expected to have a strong impact on the evolution is the inclination between the angular
momentum of the dwarf galaxy disk and its orbital angular momentum. However, in the studies mentioned above
only a narrow range of inclinations was studied in detail, namely those with values $i = 0^\circ$,
$45^\circ$ and $90^\circ$. Because of this range, and the way the properties of the dwarf were measured, no
clear evidence for the dependence on this parameter was found. On the other hand, the difference between the
prograde and retrograde galaxy encounters has been recognized for a long time and known to lead to very different
outcomes (e.g. Holmberg 1941; H\'{e}non 1970; Kozlov et al. 1972; Toomre \& Toomre 1972;
Keenan \& Innanen 1975).
The issue has been recently addressed again by D'Onghia et al. (2009, 2010)
using an improved version of the impulse approximation applied to rotating systems.
Although this approximation is not directly applicable to our simulations, we attempt a comparison between these results
and the numerical ones.

\begin{table*}
\begin{center}
\caption{Initial conditions for the simulations}
\begin{tabular}{lrrrcl}
\hline
\hline
Simulation    &  $L_{\rm X}$ & $L_{\rm Y}$ & $L_{\rm Z}$  & Inclination & Line color/type \\
              &              &             &              &    (deg)    &           \\
\hline
\ \ \ \ I0 &  0.0 &  0.0   &   1.0  &\ \ 0  & red/solid    \\

\ \ \ \ I90 &  0.0 & $-1.0$ &   0.0  &\ 90  & blue/short-dashed  \\

\ \ \ \ I180 &  0.0 &  0.0   & $-1.0$ & 180 & green/dotted  \\

\ \ \ \ I270 &  0.0 &  1.0   &  0.0   & 270 & cyan/long-dashed \\

\hline
\label{initial}
\end{tabular}
\end{center}
\end{table*}

In this paper we aim at clarifying the issue of the dependence of the results
of tidal encounters between dwarfs and their hosts on the inclination of the dwarf's disk. For this
purpose we performed four simulations of tidal evolution of a dwarf galaxy orbiting a Milky Way-like host
with disk inclinations $i = 0^\circ$, $90^\circ$, $180^\circ$ and $270^\circ$. The angles of $0^\circ$ and
$180^\circ$ correspond to exactly prograde and exactly retrograde orientations of the dwarf's disk. We also measured
the properties of the dwarf galaxy in a different way that enables clear comparisons between different runs.
Preliminary results of this study, using lower resolution simulations, were discussed in {\L}okas \& Semczuk (2014).

The paper is organized as follows. In section 2 we present the simulations used in this study. In section 3 we discuss
the properties of the dwarf galaxies as they evolve in time, focusing on their kinematics, morphology and density
profiles.
Section 4 compares the results of the simulations to the
predictions of semi-analytic models. The conclusions follow in section 5.

\hspace{0.2in}
\section{The simulations}

The simulations used in this study were similar to those described in detail in {\L}okas et al. (2014a). Here we
therefore provide only a short summary.
The initial conditions for the simulations consisted of $N$-body realizations of two galaxies:
the Milky Way-like host and the dwarf galaxy, generated via procedures described in Widrow \& Dubinski (2005)
and Widrow et al. (2008). Both galaxies contained exponential disks embedded in
NFW (Navarro et al. 1997) dark matter haloes, each made of
$10^6$ particles ($4 \times 10^6$ total). We note that the results presented here differ only slightly (are less noisy)
from those of lower resolution simulations in {\L}okas \& Semczuk (2014) where a smaller number of particles
was used (by a factor of five). We are therefore confident that our present resolution is sufficient
to grasp all the essential features of the evolution.

The dwarf galaxy model had a dark
halo of mass $M_{\rm h} = 10^9$ M$_{\odot}$ and concentration $c=20$. Its disk had a
mass $M_{\rm d} = 2 \times 10^7$ M$_{\odot}$, an exponential scale-length $R_{\rm d} = 0.41$ kpc and thickness
$z_{\rm d}/R_{\rm d} = 0.2$. The model is stable against formation of the bar in isolation for the
time scales of interest here.
The host galaxy was similar to the model MWb of Widrow \& Dubinski (2005). It had a dark matter halo of mass
$M_{\rm H} = 7.7 \times 10^{11}$ M$_{\odot}$ and concentration $c=27$. The disk of the host had a mass $M_{\rm D}
= 3.4 \times 10^{10}$ M$_{\odot}$, the scale-length $R_{\rm D} = 2.82$ kpc and thickness $z_{\rm D} = 0.44$ kpc.
The disk was also stable against bar formation to avoid strong variations of the host potential in time. The disk
of the Milky Way was coplanar with the orbit of the dwarf. Although this may seem contrary to observational
constraints where most of satellite orbits are found to be polar (e.g. Pawlowski \& Kroupa 2014),
the choice was motivated by the necessity to
avoid any additional variability which may be due to the passages through the plane of the Milky Way disk. However,
we have performed an additional simulation to verify that for the orbits used here the evolution of the dwarf depends
very weakly on the orientation of the orbit with respect to the Milky Way disk.

The dwarf galaxy was initially placed at an apocenter of a typical, eccentric orbit around the Milky Way
with apo- to pericenter distance ratio of $r_{\rm apo}/r_{\rm peri} = 120/25$ kpc. The initial position was
at the coordinates $(X,Y,Z) = (-120,0,0)$ kpc of the simulation box and the velocity vector of the dwarf was toward the
negative $Y$ direction. We performed four simulations
with different dwarf disk orientations with respect to the orbit: two coplanar with the orbit (prograde and retrograde)
and two perpendicular to the orbit with angular momenta in the same and opposite direction to the dwarf's orbital
velocity. The different initial conditions, in particular the components of the unit angular momentum vector, are
listed in Table~\ref{initial}. We will refer to the simulations by names indicating the initial inclination of the
disk, I0, I90, I180 and I270, where the inclination is measured as the rotation angle around the $X$ axis of the
simulation box.

The evolution of the system in each simulation was followed for 10 Gyr using the GADGET-2 $N$-body code
(Springel et al. 2001; Springel 2005) with outputs saved every 0.05 Gyr.
The adopted softening scales were $\epsilon_{\rm d} = 0.02$ kpc and
$\epsilon_{\rm h} = 0.06$ kpc for the disk and halo of the dwarf while $\epsilon_{\rm D} = 0.05$ kpc and
$\epsilon_{\rm H} = 2$ kpc for the disk and halo of the host, respectively.

\hspace{0.2in}
\section{Evolution of the dwarfs}

In this section we look at the inner properties of the dwarf galaxies as they are transformed by the tidal forces from
the Milky Way. All measurements discussed below were made for stars (and dark matter particles) within the radius
of 0.5 kpc from the center of the dwarf.

\subsection{Mass content}

\begin{figure}
\begin{center}
    \leavevmode
    \epsfxsize=8cm
    \epsfbox[0 0 186 190]{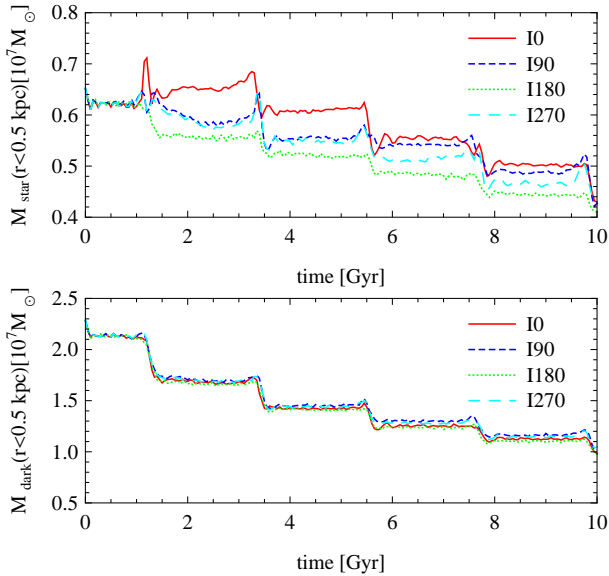}
\end{center}
\caption{The evolution of the stellar (upper panel) and dark (lower panel) mass of the dwarf galaxy
	enclosed within the radius of 0.5 kpc. Different lines correspond to different initial inclination
	of the dwarf galaxy disk.}
\label{compmass}
\end{figure}

We begin the analysis by measuring the mass inside this radius.
Figure~\ref{compmass} compares the mass of stars (upper panel) and dark matter (lower panel) in the four simulations.
As expected, the dark mass content decreases systematically, most significantly at pericenters that occur at
$t = 1.2$, 3.3, 5.5, 7.6, 9.7 Gyr from the start of the simulation and there is no dependence on the inclination
because the halos were spherical and isotropic in all cases.

On the other hand, for the stellar component (upper panel)
there is a significant difference between dwarfs with varying initial inclination. However, the dependence may seem
surprising because we expect the prograde disk to be more stripped, while the opposite is seen in the Figure: the
stellar mass for the I0 simulation is even increased after the first pericenter passage, while it is decreased in the
remaining cases. While the tidal stripping is indeed stronger in the outer parts, in the inner region we probe by
this measurement, the stellar content increases as a result of the significant change in the structure
of the stellar component due to the formation of a tidally induced bar.

\subsection{Kinematics}

The main criteria usually applied in order to verify if a dwarf galaxy transformed from a disky object to a
dSph are based on kinematics and shape of the stellar component. A dSph galaxy is supposed to be characterized
by the dominance of random motions of the stars over the amount of rotation and its shape should be
sufficiently close to spherical. In this and the following subsection we look in detail at the evolution of
these quantities.

In order to measure these properties, for each simulation output we determine the directions of the principal
axes of the stellar component from stars within the radius of 0.5 kpc using the inertia tensor and rotate
these stars so that the new coordinate system is aligned with the principal axes (the $x$ axis is along the longest,
the $y$ axis along the intermediate and the $z$ axis along the shortest axis of the stellar component).
We then introduce a standard spherical coordinate system such that $\phi$ measures the angle around the $z$
axis and $\theta$ the angle from the $z$ axis towards the $xy$ plane.

\begin{figure}
\begin{center}
    \leavevmode
    \epsfxsize=8cm
    \epsfbox[0 0 186 279]{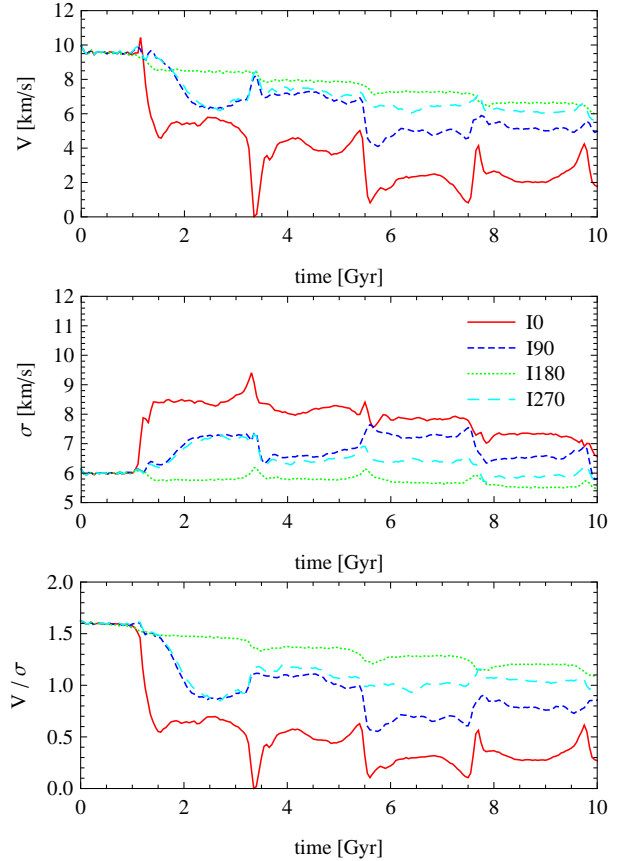}
\end{center}
\caption{The evolution of the mean rotation velocity (upper panel), the velocity dispersion (middle panel)
and the ratio of the two (lower panel) as a function of time.}
\label{compvsig}
\end{figure}

The kinematic properties were estimated using these coordinates. In the top panel of Figure~\ref{compvsig}
we plot the rotation around the shortest axis $V=V_\phi$ as a function of time. We note that there is no other
significant streaming motion along the other spherical coordinates or around the two other principal axes (see
{\L}okas et al. 2014b for a brief discussion of this issue). Clearly, the rotation is decreasing most strongly
for the exactly prograde inclination of the disk (I0), a significantly smaller decrease is seen for the perpendicular
orientations (I90 and I270) and the effect is the weakest for the retrograde case (I180). Note that the decrease of
rotation is not steadily monotonic even in the exactly prograde case (I0). This is due to the tidal torques acting
on the bar at pericenter passages that can speed up or slow down the bar depending on its particular orientation at
this moment (see {\L}okas et al. 2014a for details).

The behavior of the velocity dispersion is exactly the opposite. In the middle panel of Figure~\ref{compvsig}
we show the 1D velocity dispersion calculated as
$\sigma = [(\sigma_r^2 + \sigma_{\theta}^2 +  \sigma_{\phi}^2)/3]^{1/2}$. The increase of $\sigma$ at the first
pericenter passage is strongest for the I0 case, intermediate for I90 and I270 and for I180 $\sigma$ remains
constant in time or even slightly decreases due to mass loss. The ratio $V/\sigma$ shown in the lower panel of
Figure~\ref{compvsig} decreases for all simulations, but reaches a value significantly below unity only for the
prograde case. For the three remaining cases a substantial amount of rotation is retained, although the hierarchy
of lower $V/\sigma$ for more prograde cases is preserved.

\begin{figure}
\begin{center}
    \leavevmode
    \epsfxsize=8cm
    \epsfbox[0 0 186 358]{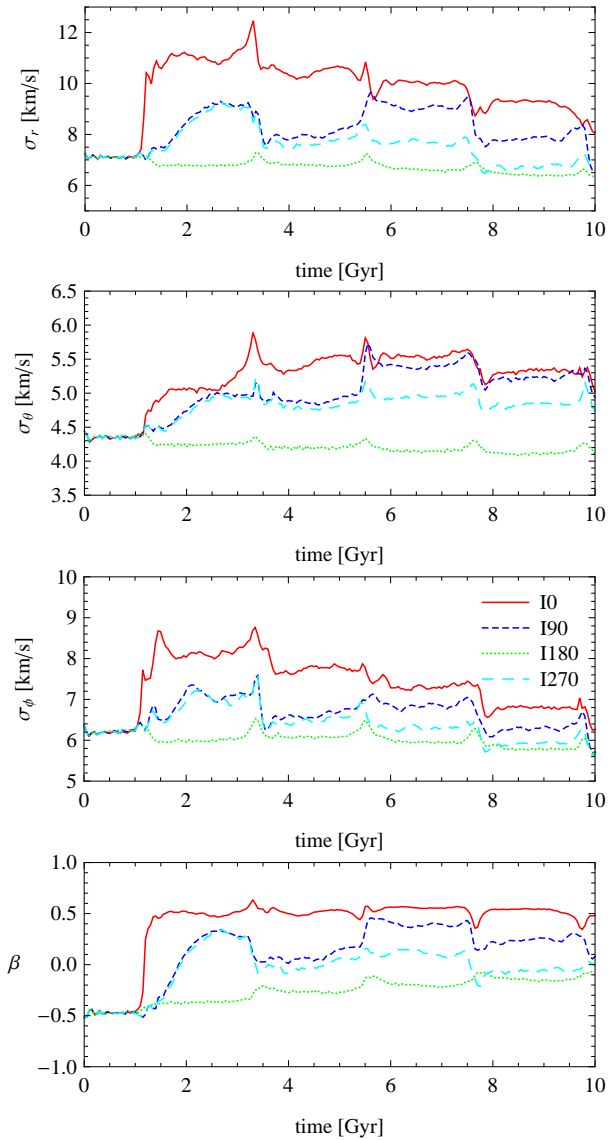}
\end{center}
\caption{The evolution of the velocity dispersions of the stars in spherical coordinates $\sigma_r$ (upper panel),
$\sigma_\theta$ (second panel) and $\sigma_\phi$ (third panel). The lower panel shows the evolution of the
anisotropy parameter $\beta$.}
\label{compsigma}
\end{figure}

In Figure~\ref{compsigma} we show the evolution of the different velocity dispersions $\sigma_r$ (upper panel),
$\sigma_\theta$ (second panel) and $\sigma_\phi$ (third panel) as a function of time. In all cases the increase
of a given dispersion at the first pericenter passage and its level at later times is highest for the I0 case,
intermediate for I90 and I270 and non-existent for I180. In addition, this increase is the most abrupt for the
prograde dwarf, while for the intermediate inclinations I90 and I270 the increase occurs much more slowly in
time and takes about half the orbital period between the first and second pericenter passage. Interestingly,
significant increase is seen in all dispersions in spite of the fact that, due to the formation of the bar, one
could expect the radial $\sigma_r$ to increase much more significantly.

\begin{figure}
\begin{center}
    \leavevmode
    \epsfxsize=8cm
    \epsfbox[0 0 186 358]{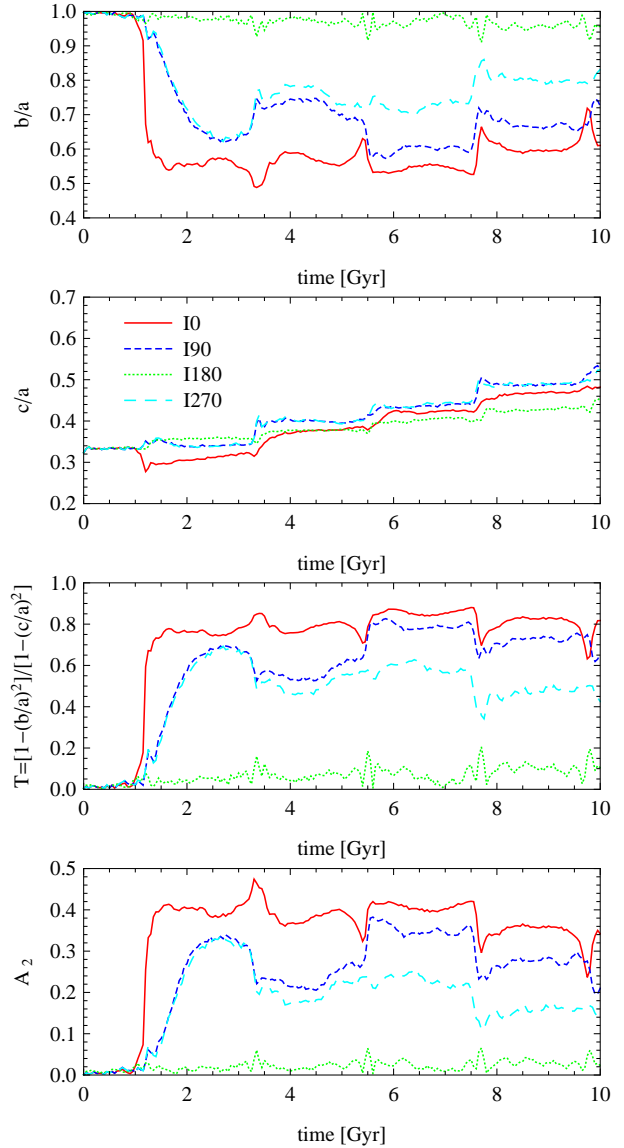}
\end{center}
\caption{The evolution of the shape of the stellar component in time. The panels from top show the axis ratios
$b/a$, $c/a$, the triaxiality parameter $T$ and the bar mode $A_2$.}
\label{compshape}
\end{figure}

We end the analysis of the kinematics by plotting in the lower panel of Figure~\ref{compsigma} the value of the
anisotropy parameter $\beta = 1 - \sigma_t^2/(2 \sigma_r^2)$ where $\sigma_t^2 = \sigma_\theta^2 + \sigma_\phi^2
+V_\phi^2$ is the tangential second moment including rotation. A systematic dependence of this parameter on the
inclination of the disk is also present: the stellar orbits are most radial for the I0 case with $\beta$ almost
constantly at the level of 0.5, corresponding to mildly radial orbits of the stars, characteristic of the bar.
For the intermediate cases I90 and I270 the $\beta$ values stay between 0 and 0.4,
while for the retrograde case I180 $\beta$ remains negative due to the dominant presence of rotation.
Interestingly, at the end of the evolution all three non-prograde cases I90, I180 and I270 have almost isotropic
orbits ($\beta=0$) although they reach this special value via different evolutionary paths.

\subsection{Shapes}

Figure~\ref{compshape} illustrates the evolution of the shape of the stellar component of the dwarfs in time.
In the first and second panels from top we plot the axis ratios $b/a$ (intermediate to longest) and $c/a$
(shortest to longest). The thickening of the dwarf, as quantified by the increasing value of $c/a$ is
similar in all cases, although for the prograde case $c/a$ decreases after the first pericenter due to the formation of
the bar. Much more significant differences are seen in the evolution of $b/a$. This value decreases most
strongly for the prograde I0 case signifying the prolate shape characteristic of the bar. The intermediate
inclinations I90 and I270 lead to less prolate shapes, while I180 remains disky for the whole evolution, as
indicated by $b/a$ remaining constantly close to unity.

The shape can be also quantified in terms of the triaxiality parameter $T = [1-(b/a)^2]/[1-(c/a)^2]$ which is shown
in the third panel of Figure~\ref{compshape}. The values of
the parameter $T>2/3$ at all times confirm the prolate shape due to the bar in the case of I0. For the intermediate
cases I90 and I270 we get $1/3 < T < 2/3$ indicating a triaxial shape. For the retrograde case I180 the parameter
remains low, $T < 0.2$, at all times as is characteristic of a disk.

\begin{figure}
\begin{center}
    \leavevmode
    \epsfxsize=4.1cm
    \epsfbox[0 0 185 200]{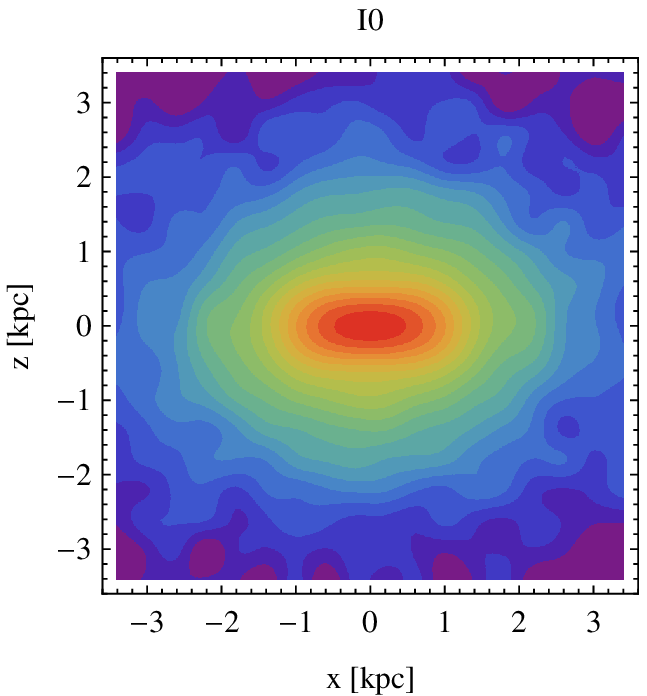}
\leavevmode
    \epsfxsize=4.1cm
    \epsfbox[0 0 185 200]{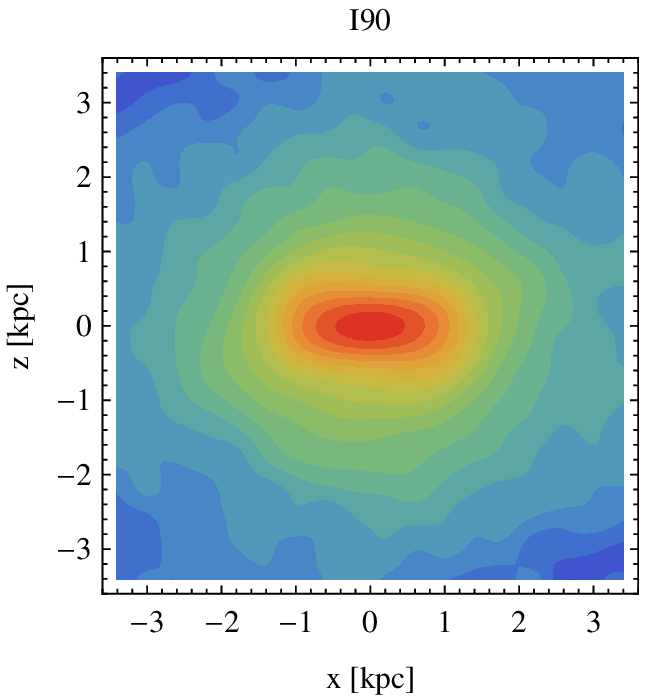}
\leavevmode
    \epsfxsize=4.1cm
    \epsfbox[0 0 185 200]{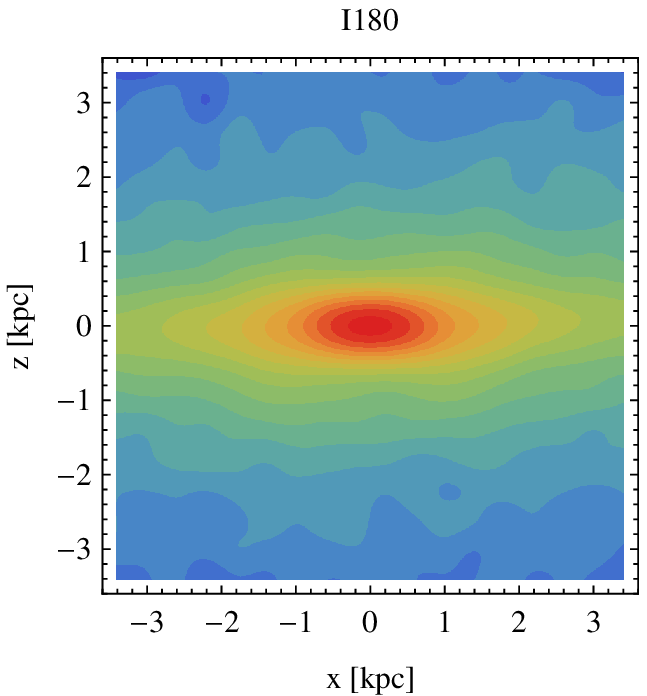}
\leavevmode
    \epsfxsize=4.1cm
    \epsfbox[0 0 185 200]{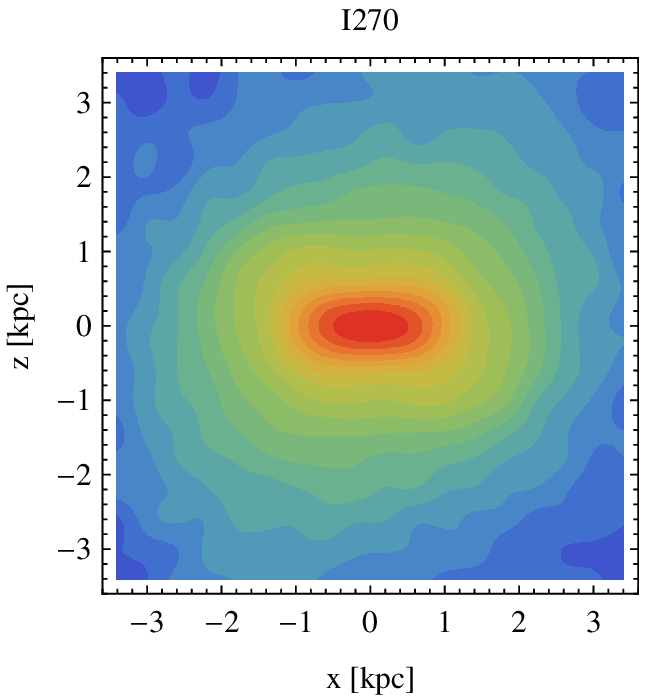}
\end{center}
\caption{Surface density distributions of the stars in the dwarfs at the last apocenter
($t=8.65$ Gyr) seen along the intermediate ($y$)
axis of the stellar component. The surface density
measurements were normalized to the maximum value $\Sigma_{\rm max} = 4.8 \times 10^5$ stars kpc$^{-2}$ occurring
for I180. Contours are equally spaced in $\log \Sigma$ with $\Delta \log \Sigma = 0.05$.}
\label{surdenrot}
\end{figure}

The presence of a bar is usually detected by measuring the bar mode
$A_2$ of the Fourier decomposition of the stars projected along the shortest axis of the stellar distribution
(see a more detailed discussion in {\L}okas et al. 2014a).
Usually, $A_2 > 0.3$ is considered as high enough to be interpreted as a bar. As we can see in the lower panel
of Figure~\ref{compshape} this is always the case for simulation I0 after the first pericenter passage and also
for some significant periods of time for the intermediate cases I90 and I270, which means that the bar also forms
there, but it is much weaker. The exactly retrograde disk does not
form a bar as its $A_2 < 0.06$ at all times. Slight temporary increases of this value are due to stretching of
the dwarf at the pericenters.

These measurements are confirmed by the maps of the surface density distribution of the stars in the four dwarfs
plotted in Figure~\ref{surdenrot}. The distributions are shown in projection along the intermediate axis of the
stellar component so in their most non-spherical appearance. The snapshots were selected for the time $t=8.65$ Gyr
after the start of the simulations, corresponding to the last apocenter passage. In all cases, except for the
retrograde one I180, the remnant of the bar formed after the first pericenter passage is still visible in the
inner parts. In addition, the distribution of the stars in the prograde case I0 is much less diffuse in the outer
parts of the maps. In spite
of the fact that this dwarf evolves most strongly, its neighborhood is not uniformly filled with stripped debris
because the lost stars form well-defined, narrow tidal streams.

\subsection{Density profiles}

\begin{figure}
\begin{center}
    \leavevmode
    \epsfxsize=7cm
    \epsfbox[0 0 174 174]{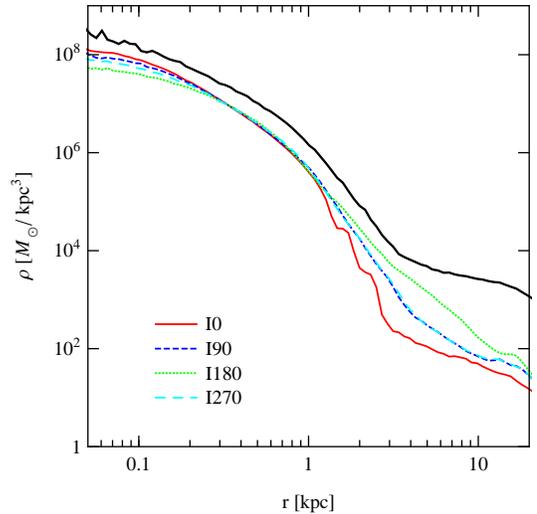}
\end{center}
\caption{Comparison of stellar density profiles of the dwarfs
at the fourth apocenter (6.6 Gyr after the start of the simulations, colored lines).
For completeness we also show the dark matter profile (black line) which is similar for all simulations.}
\label{compprofiles}
\end{figure}

The dependence of the properties of the dwarfs on the initial inclination of the disk also manifests itself in
the evolved density profiles. In Figure~\ref{compprofiles} we show examples of the stellar density profiles
(different colors) and the dark matter
density profile (black line) for different simulations considered here, measured at the fourth apocenter.
The transition from the bound component to the tidal tails is visible as the break in the slope of the density
profiles. This transition is however only well defined
for the exactly prograde case I0 (and the dark matter profile). In this case the transition from the steeper
to the shallower profile (where the slope is around
$r^{-4}$) occurs at around 3 kpc. For other simulations no such clear break radius is seen.

As discussed in {\L}okas et al. (2013) using similar (but only mildly prograde) simulation setups, the break radii
can be interpreted as the tidal radii. In this case, the dependence on the orbit of the star within the satellite
is expected (Keenan \& Innanen 1975; Read et al. 2006) and we will attempt a detailed comparison in a
follow-up paper. Since the
stellar density profiles of our simulated dwarfs do not show clear signatures of the
break radius (except for the exactly prograde case) here we propose a comparison in terms of density.
The stellar profiles shown in Figure~\ref{compprofiles} demonstrate clear hierarchy:
at the outer radii (larger than 1 kpc) the stellar density profile of I180 (green line) is above all the other
profiles, the one of I0 (red lines) is the lowest, and the ones of I90 and I270 fall exactly on top of each other
and between the other two. This means that we can quantify the amount of tidal stripping in these different cases by
measuring the density of the stars, rather than the break radius.

\begin{figure}
\begin{center}
    \leavevmode
    \epsfxsize=8cm
    \epsfbox[0 0 186 178]{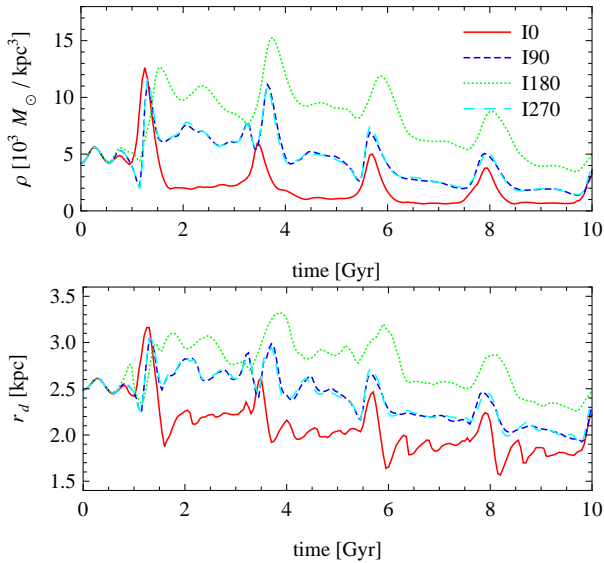}
\end{center}
\caption{Upper panel: the density of stars in the shell of radii 2 kpc $< r <$ 4 kpc
as a function of time. Lower panel: the radius at which the density of stars drops below $10^4 $M$_{\odot}/$kpc$^3$
as a function of time.}
\label{comptidal}
\end{figure}

To do so, we calculated the mean density of stars in the shells of radii 2 kpc $< r <$ 4 kpc
as a function of time for different simulations. The results are shown in the upper panel of Figure~\ref{comptidal}.
In the lower panel we plot the radius $r_d$ at which the density of stars drops below $10^4 $M$_{\odot}/$kpc$^3$
as a function of time. In both plots there is a clear systematic difference between the measurements for different
initial disk orientations: the stars are stripped more effectively on prograde orbits as demonstrated by the lower
outer densities and smaller radii where the density drops to a fixed value.
Although the values for the measurements were chosen in an arbitrary way, we expect the results to be similar if
these parameters are slightly modified. They clearly confirm that the amount of tidal stripping depends
very strongly on the initial inclination of the dwarf's disk.

\section{Comparison with semi-analytic predictions}

To describe the encounter between a disky dwarf galaxy and the Milky Way we use the impulse approximation
as discussed in Binney \& Tremaine (1987) and D'Onghia et al. (2010). According to this approximation the
$k$th component of the acceleration of each dwarf's star with respect to the center of mass is given by
\begin{equation}      \label{acceleration}
	\dot{v}_k=-\sum_j \frac{\partial^2\psi}{\partial x_k \partial x_j}\at[\bigg]{\boldsymbol{x}=0}x_j,
\end{equation}
where $\psi$ is the Milky Way's (i.e. perturber's) potential and $x_i$ are Cartesian coordinates. We integrate equation
(\ref{acceleration}) over a finite time period to obtain velocity increments that can be compared with
increments measured from simulations
\begin{equation}      \label{deltav}
	\Delta v_k=\int\limits_0^{\Delta t}\dot{v}_k\; {\rm d} t.
\end{equation}

\begin{figure*}
\begin{center}
    \leavevmode
    \epsfxsize=17cm
    \epsfbox[0 0 610 650]{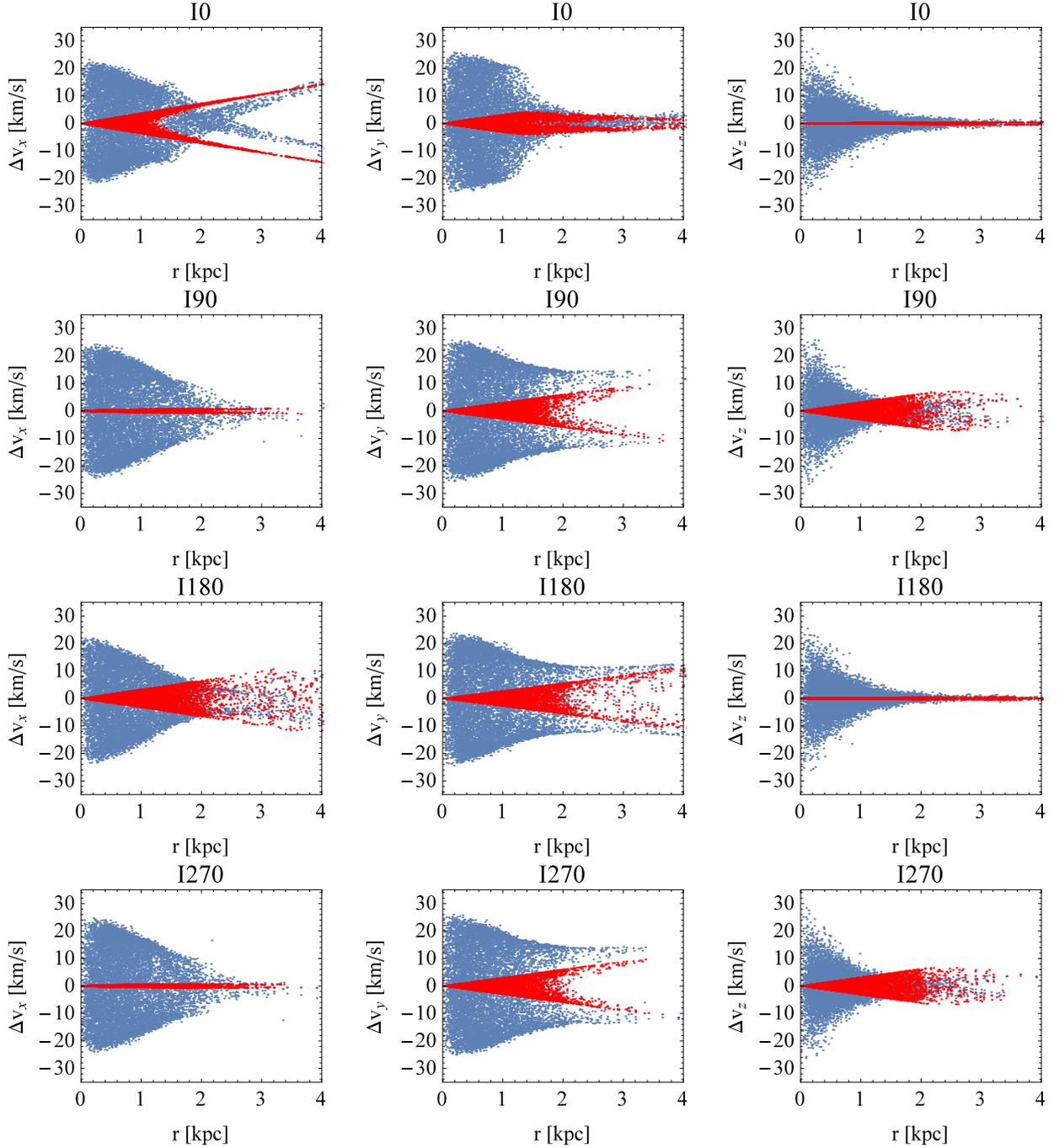}
\end{center}
\caption{The components of velocity increments along the $x$, $y$ and $z$ axis for simulations with different
initial inclination of the dwarf's disk. The blue
dots indicate the values measured for a subsample of stars in the simulation. Red dots show the corresponding
semi-analytic predictions.}
\label{velocities}
\end{figure*}

We work in a frame centered on the center of the dwarf's mass, the dwarf's disk lies in $xy$ plane and
position of the Milky Way at the pericenter is $(b_x,0,b_z)$.
In this frame the trajectory of each star originating from the dwarf is
\begin{equation}       \label{trajectory}
	\boldsymbol{x}=[r\cos(\Omega t +\phi_0), r\sin(\Omega t +\phi_0), 0],
\end{equation}
where $r$ is the radius, $\Omega$ is the angular velocity and $\phi_0$ is the initial azimuthal angle
of the star. The trajectory of Milky Way is given by
\begin{equation}       \label{trajectoryMW}
	\boldsymbol{X}=\boldsymbol{V} t+\boldsymbol{X_0},
\end{equation}
where $\boldsymbol{V}$ and $\boldsymbol{X_0}$ are constant vectors, fitted to mimic the perturber's trajectory
from simulations as a straight line during a given time period.
Note that $\boldsymbol{V}$ and $\boldsymbol{X_0}$ are different for each of our simulations as
they depend on the inclination $i$. This dependence can be found by rotating the
trajectory for the prograde case with matrix $\hat{A}$ defined by Euler angles, in order to obtain
trajectories in other cases. One of the Euler angles is the inclination $i$ and other two depend
on the perturber's orbit.

To approximate the gravitational potential of the Milky Way we sum the potential from its stars and
the dark matter halo.
The first part is represented as a point-mass potential, while the second is given by the NFW profile
\begin{equation}       \label{potentialMW}
	\psi=-\frac{G M_{\rm D}}{|\boldsymbol{x}-\boldsymbol{X}|}-g\frac{G M_{\rm H}\ln(1+
	c |\boldsymbol{x}-\boldsymbol{X}|/r_{\rm{v}})}{|\boldsymbol{x}-\boldsymbol{X}|},
\end{equation}
where $M_{\rm D}$ is the mass of the Milky Way disk, $M_{\rm H}$ is the virial mass of its halo, $r_{\rm v}$ is
the virial radius, $c$ is the concentration parameter and $g=[\ln(1+c)-c/(1+c)]^{-1}$ (see {\L}okas \& Mamon 2001).
Substituting equations (\ref{acceleration}), (\ref{trajectory}), (\ref{trajectoryMW}) and (\ref{potentialMW})
into (\ref{deltav}) we obtain formulae which can be numerically integrated to get
velocity increments.

\begin{figure*}
\begin{center}
    \leavevmode
    \epsfxsize=7cm
    \epsfbox[0 0 380 400]{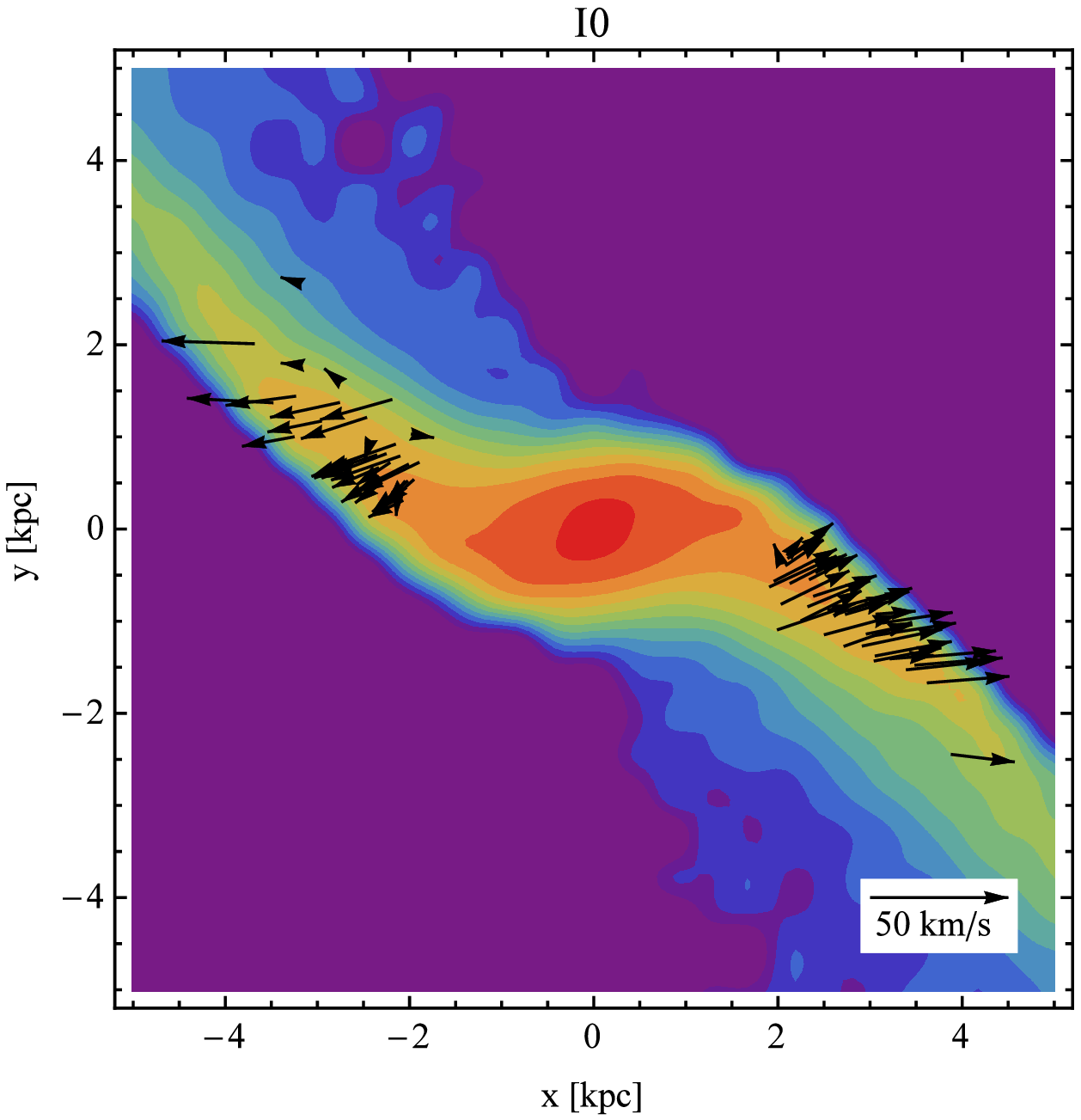}
\leavevmode
    \epsfxsize=7cm
    \epsfbox[0 0 380 400]{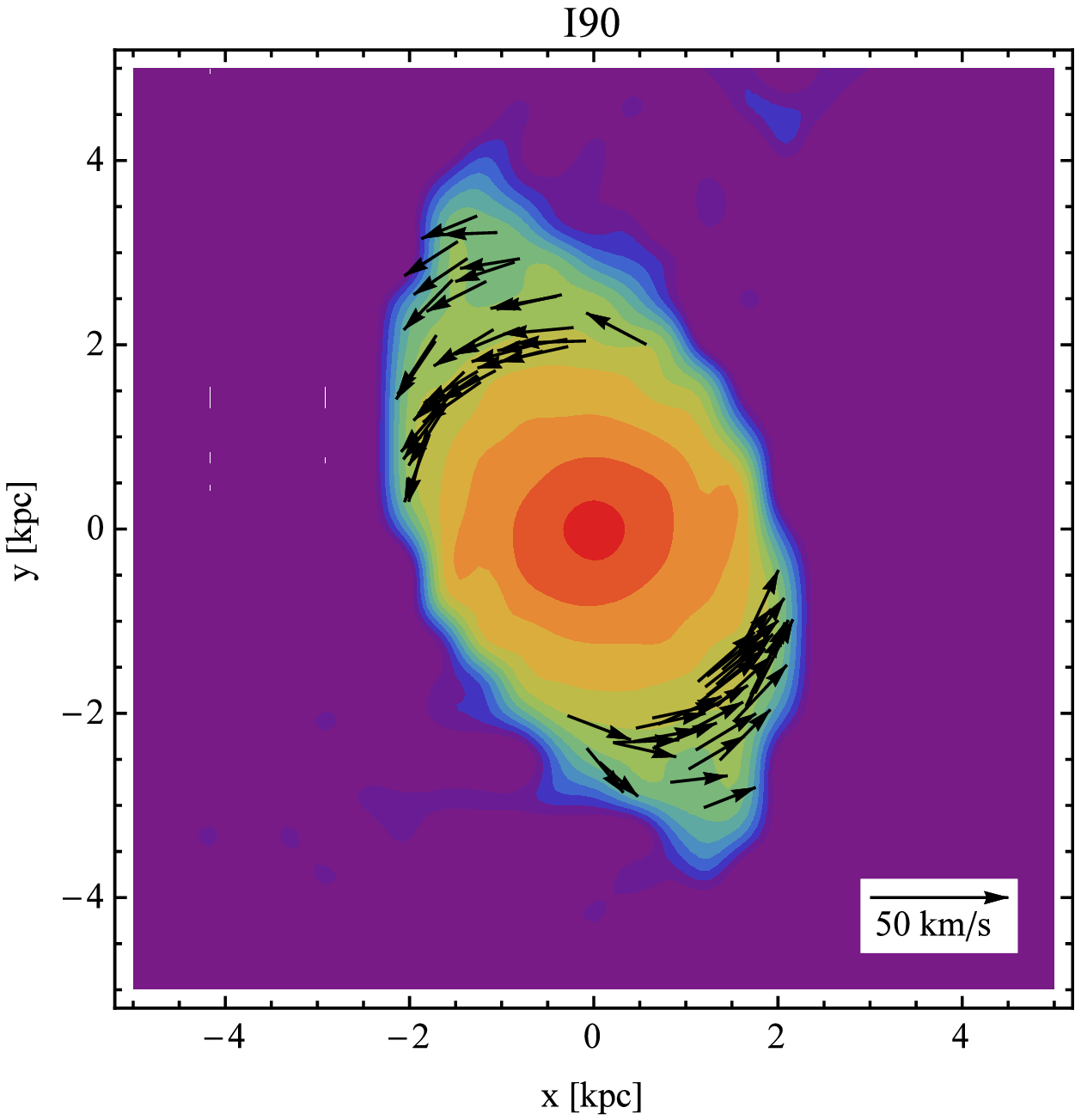}
\leavevmode
    \epsfxsize=7cm
    \epsfbox[0 0 380 400]{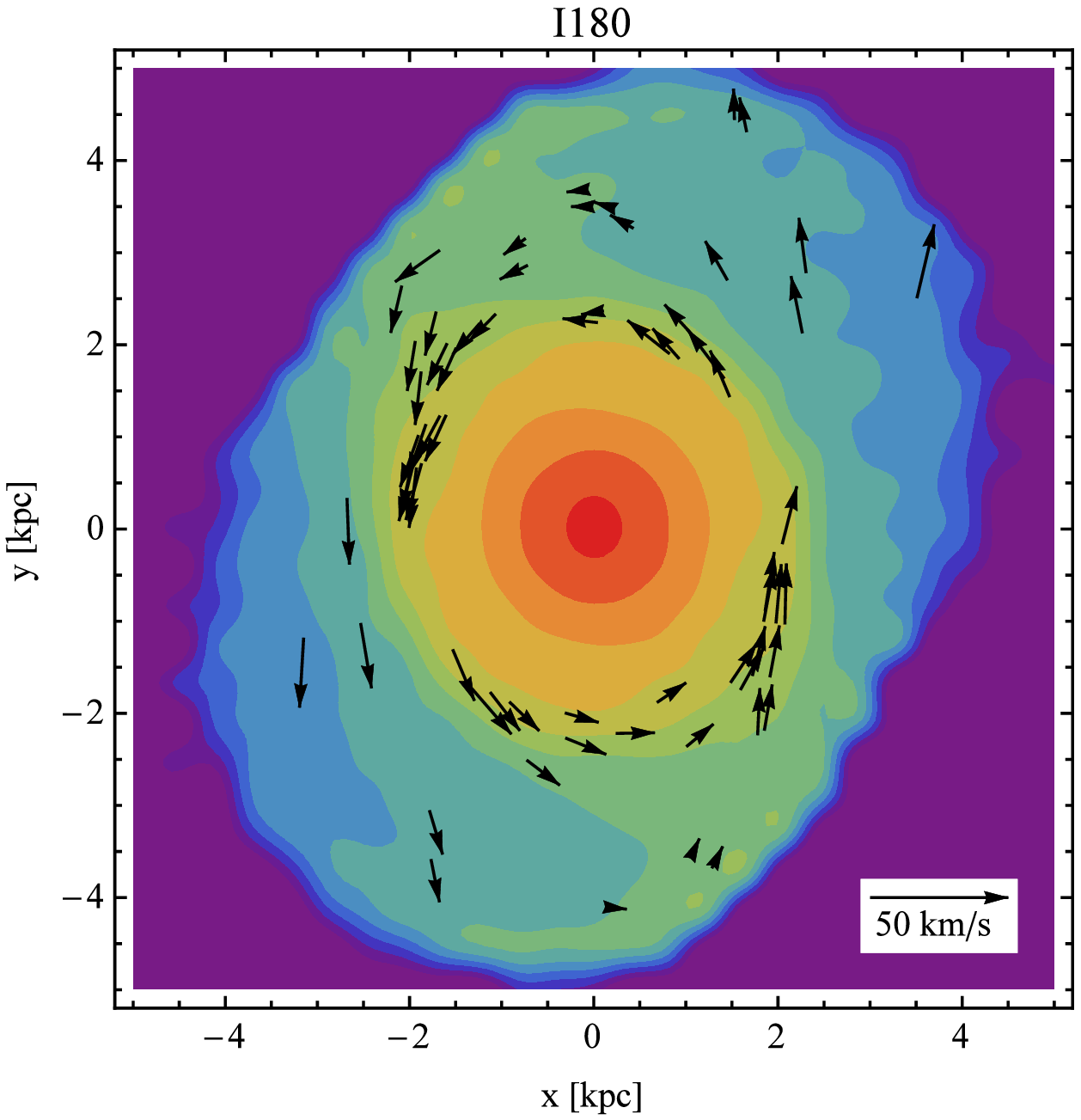}
\leavevmode
    \epsfxsize=7cm
    \epsfbox[0 0 380 400]{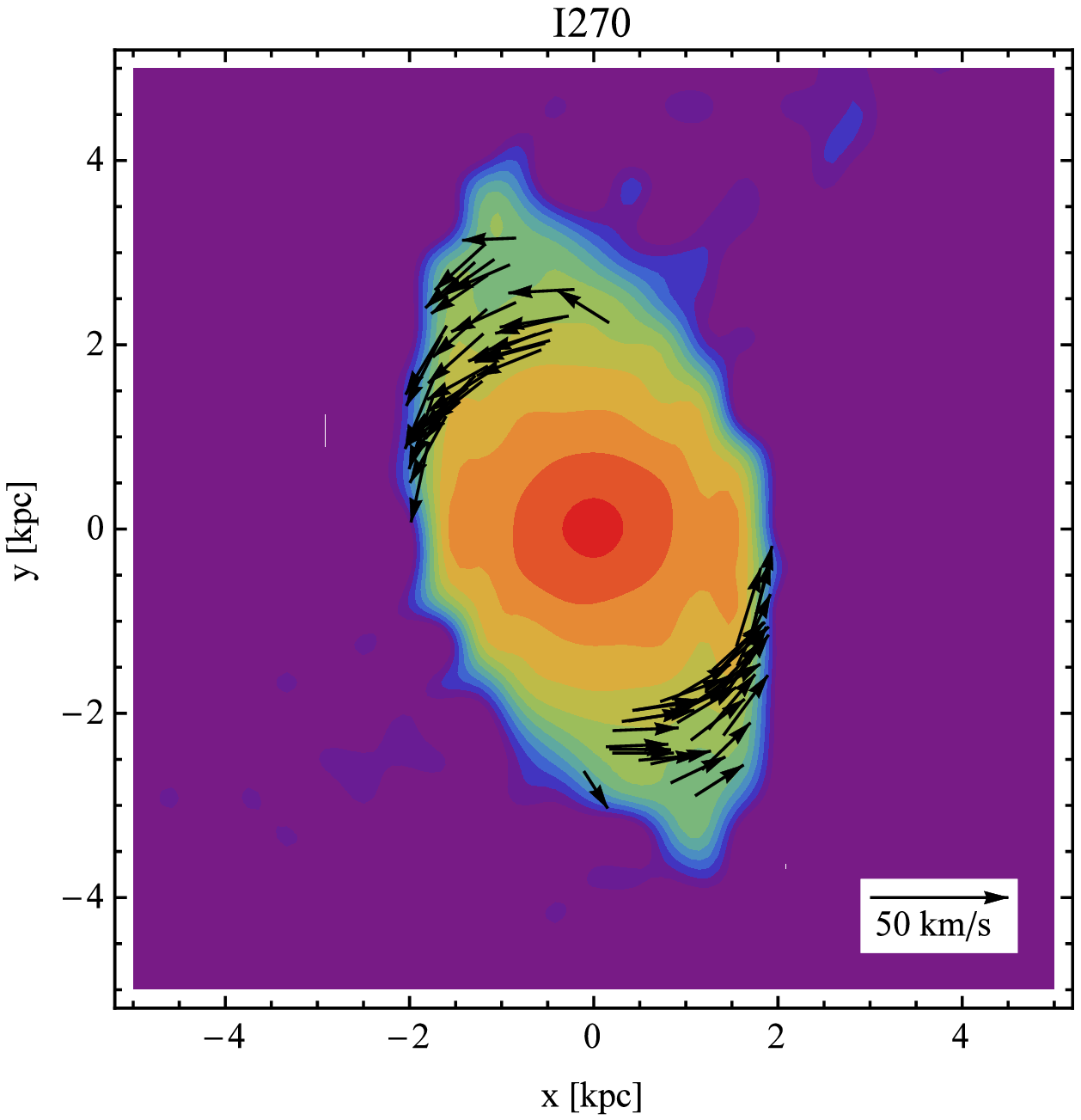}
\end{center}
\caption{Surface density distributions of the stars in the dwarfs at the first pericenter
($t=1.15$ Gyr) in projection onto the initial disk plane. Arrows indicate the
velocities of the randomly selected subsample of 100 stars outside the radius of 2 kpc.}
\label{pericenter}
\end{figure*}

In order to calculate the velocity increments described above we estimate the angular velocity in the dwarf as
$\Omega=|\boldsymbol{v}|/r$. We calculate integrals of equation (\ref{deltav}) over a small period of time,
so that the assumptions concerning the trajectories are valid. In our simulations the outputs were saved every 0.05 Gyr
and we choose to integrate over this time to compare our predictions with velocity increments occurring in
the simulations when the tidal force is the strongest, i.e. between the output preceding the first pericenter
and the one as close as possible to this pericenter. We check how the distribution of increments changes
with the distance from the center of the dwarf. The results for all simulations are summarized in
Figure~\ref{velocities}.

The red points in Figure~\ref{velocities} represent the values predicted by the impulse approximation,
while the blue points correspond
to the values measured from the simulations. At radii smaller than 2 kpc the semi-analytical predictions do not
reproduce the distribution of velocity increments because at these radii the velocity changes are dominated by the
gravitational potential of the dwarf which was not included in the predictions.
However, for radii greater than 2 kpc the agreement between the full simulations and semi-analytic predictions
for this short time period is very good. In particular, in the upper left panel of Figure~\ref{velocities}
for simulation I0 we find two very well-defined
branches corresponding to stars on different sides of the dwarf. Branches from simulations
are not exactly symmetric with respect to zero while the branches from theoretical predictions are.
The difference is due to the fact that the analytic predictions only take into account the lowest order terms.

For simulations I270 the increments are almost identical as for I90, as expected due to symmetry of the
two configurations with respect to the orbital plane.
In some of the panels in Figure~\ref{velocities} the velocity increments for radii
larger than 2 kpc are approximately zero. However,
the negligible values are consistently obtained both from the simulations and the semi-analytic calculations.

We further illustrate these results in Figure~\ref{pericenter} where we plot the surface density maps of
the stellar component
and the velocity vectors for a random sample of a hundred stars at radii larger than 2 kpc.
The plots show the dwarfs at the first pericenter passage, i.e. after they have been affected by a tidal impulse
from the Milky Way for the first time. The coordinate system is as defined above, with $xy$ coordinates in the plane
of the dwarf's disk.

The comparison of the upper left panel of Figure~\ref{pericenter}, corresponding to simulation I0, to the other three
confirms that for the prograde encounter the effect of
the tidal force in the strongest: the dwarf galaxy disk is already strongly distorted toward a bar-like shape and
two tidal arms are formed. There are significant increments of velocity along the $x$ axis.
Small increments of velocities are also visible along the $y$ axis in the right panels of Figure~\ref{pericenter}
corresponding to simulations I90 and I270.
The effect of the tidal force is weakest for the retrograde case (lower left panel of Figure~\ref{pericenter})
where the dwarf's initial disk is affected very little.

As discussed by D'Onghia et al. (2010) strongest tidal interactions occur when the intrinsic angular velocities of
stars in the dwarf's disk are comparable to the angular velocity of the satellite on its orbit.
This condition can be written as
\begin{equation}         \label{resonance}
	\Omega_{\rm disk} \simeq \Omega_{\rm orb}
\end{equation}
and we can define the resonance parameter
\begin{equation}          \label{resparameter}
	\alpha=\frac{|\Omega|}{\Omega_{\rm orb}},
\end{equation}
that should be of the order of unity for the strongest, resonant response.

\begin{figure}
\begin{center}
    \leavevmode
    \epsfxsize=8cm
    \epsfbox[0 0 186 94]{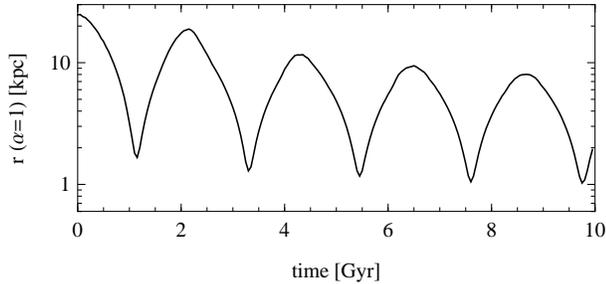}
\end{center}
\caption{The evolution of the radius at which the resonance parameter $\alpha=1$ in time. Only results for
simulation I0 are shown because the measurements yield similar values for the remaining simulations.}
\label{alpha}
\end{figure}

To demonstrate the resonant nature of the tidal effects in our simulations we measured $\alpha (r)$ for each
simulation output and found radii from the center of the dwarf at which $\alpha =1$. The time dependence of
this radius for simulation I0 is shown in Figure~\ref{alpha}. We can see that the variability
of this radius reflects the varying orbital velocity $\Omega_{\rm orb}$
of the dwarf resulting in rather large values at apocenters
and much smaller ones at pericenters. The slow variation over time scales much larger than the orbital period is caused
by the mass loss and decreasing $\Omega$.
Whenever the dwarf
reaches the pericenter of its orbit around the Milky Way, this characteristic radius drops below 2 kpc, and even
down to 1 kpc at the later pericenters. Note that these radii are of the order of 2-3 half-light radii of the
stellar component which means that a significant fraction of stars is affected.
Comparing with Figure~\ref{velocities} showing the velocities at the first pericenter we confirm that this radius
(equal to 1.7 kpc at this time) is exactly where the tidal effects start to prevail
over the dwarf's potential.

\section{Conclusions}

In this work we extended previous studies of the efficiency of the tidal stirring mechanism to include the
dependence on the initial inclination of the dwarf galaxy disk with respect to its orbit around the Milky Way.
Our simulation setups involved a dwarf galaxy placed on a typical, eccentric orbit around a Milky Way-like host and
its evolution was followed for 10 Gyr. We considered four configurations, an exactly prograde, an exactly retrograde and
two intermediate orientations of the disk. We found the efficiency of tidal stirring to be very strongly dependent on this
inclination.

The effect of the tidal interaction turns out to be the strongest for the exactly prograde orientation of the
dwarf's disk (I0). In this case the disk transforms into a strong bar ($A_2 = 0.4$)
at the first pericenter passage and a remnant bar is
retained until the end of the evolution. Although the bar becomes weaker with time, the shape of the stellar component
is consistently prolate, but tending to a spherical toward the end of the evolution. This morphological
transformation is mirrored in the kinematics by the gradual decrease of the dwarf's rotation velocity. Already at the
first pericenter passage the rotation drops significantly and the velocity dispersion increases, mostly in the
radial direction due to the formation of the bar. During subsequent pericenter passages the rotation is further
decreased down to rather low values at the end of the simulation where $V/\sigma =0.3$.
Thus the streaming motions are
almost completely replaced by random motions of the stars. Interestingly, the radial velocity dispersion dominates
the whole time, which manifests itself in the anisotropy parameter close to $\beta = 0.5$ even at the end.

In the two cases of perpendicular orientations of the dwarf's disk with respect to the orbit (I90 and I270) the
evolution is to some extent similar to the I0 case. In these configurations the bar also forms after the first
pericenter passage, but more slowly (over a time scale of about 1 Gyr corresponding to half the orbital period)
and is significantly weaker
($A_2 = 0.3$), also in the subsequent evolution. The overall shape of the stellar component can be characterized
more as triaxial than decidedly prolate. The transition from the streaming to random motions of the stars also
happens less efficiently with $V/\sigma$ only slightly below unity at the final outputs. The anisotropy parameter
is close to zero due to the contribution of the still significant rotation.

In the exactly retrograde case (I180) no strong evolution is present: the dwarf's stellar
component does not form a bar and remains disky. The only signatures of tidal evolution in this case are the mass
loss (similar as in other cases, mostly in dark matter), small decrease of the rotation velocity,
slight evolution of the anisotropy parameter from negative toward isotropic and a non-negligible thickening of the disk.
The difference between this and the other cases is also visible in the stellar density profiles which are less
affected and do not show any clear transition from the bound component to the tidal tails.

We have interpreted these changes in the context of the resonant stripping mechanism recently discussed by
D'Onghia et al. (2009, 2010). In particular, we calculated the velocity increments the dwarf's stars should
experience at the first pericenter and compared them with the direct measurements from simulations. We find a
very good agreement between the two, confirming the interpretation that the evolution we see in the full
$N$-body treatment is indeed due to the orientation of the dwarf's disk. The resonant nature of the phenomenon
is further confirmed by the behavior of the ratio between the angular velocity of the stars in the dwarf and
the angular velocity of its orbital motion. This ratio turns out to be of the order of unity only near the
pericenters and this is indeed when the tidal effects are the strongest.

The results presented here suggest that the most important mechanism underlying the tidal evolution of
disky dwarfs orbiting a bigger galaxy is indeed of resonant nature. We propose to refer to the processes of
morphological and dynamical evolution of the dwarfs we described as
`resonant stirring' in analogy to the `resonant stripping' mechanism found by D'Onghia et al. (2009, 2010)
to increase the mass loss in similar configurations. As discussed by D'Onghia et al. (2010), the resonance
is broad, hence the name `quasi-resonant stirring' would be more appropriate. In physical terms, this resonance can be
traced to the fact that the stars with $\alpha \approx 1$ remain for an extended
period of time on the line joining the dwarf galaxy and the perturber. For these stars
the tidal force (which is strongest along this line) has the longest time to
operate which results in the largest velocity increments and the largest stirring. In the context of the tidal
radius calculations, the difference between the prograde and retrograde cases comes from the change of sign of the
Coriolis force. The relation between the two approaches remains to be investigated and we plan to address this
issue in our future work.

\section*{Acknowledgments}

This work was supported in part by PL-Grid Infrastructure, the Polish National Science Centre under
grant 2013/10/A/ST9/00023, by US National Science Foundation Grant No. PHYS-1066293 and the hospitality of
the Aspen Center for Physics. ED gratefully acknowledges the
support of the Alfred P. Sloan Foundation and of the NSF Grant No. AST-1211258 and ATP-NASA Grant No. NNX14AP53G.
MS benefited from the summer student program of the Copernicus Center. We thank L.
Widrow for providing procedures to generate $N$-body models of galaxies for initial conditions.


\begin{thebibliography}{}

\bibitem[{Binney & Tremaine}(1987)]{bt87} Binney, J., \& Tremaine, S. 1987, Galactic Dynamics (Princeton, NJ:
	Princeton Univ. Press)
\bibitem[{D'Onghia et al.}(2009)]{don09} D'Onghia, E., Besla, G., Cox, T. J., \& Hernquist, L. 2009, Nature, 460, 605
\bibitem[{D'Onghia et al.}(2010)]{don10} D'Onghia, E., Vogelsberger, M., Faucher-Giguere, C. A., \& Hernquist, L. 2010,
  ApJ, 725, 353
\bibitem[{H\'{e}non}(1970)]{he70} H\'{e}non, M. 1970, A\&A, 9, 24
\bibitem[{Holmberg}(1941)]{ho41} Holmberg, E. 1941, ApJ, 94, 385
\bibitem[{Kazantzidis et al.}(2011)]{kaz11} Kazantzidis, S., {\L}okas, E. L., Callegari, S., Mayer, L.,
	\& Moustakas, L. A. 2011, ApJ, 726, 98
\bibitem[{Keenan \& Innanen}(1975)]{ki75} Keenan, D. W., \& Innanen, K. A. 1975, AJ, 80, 290
\bibitem[{Klimentowski et al.}(2009)]{k09} Klimentowski, J., {\L}okas, E. L., Kazantzidis, S.,
	Mayer, L., \& Mamon, G. A. 2009, MNRAS, 397, 2015
\bibitem[{Kozlov et al.}(1972)]{ko72} Kozlov, N. N., Syunyaev, R. A., \& \'{E}neev, T. M. 1972, SPhD, 17, 413
\bibitem[{Lokas \& Mamon}(2001)]{lm01} {\L}okas, E. L., \& Mamon, G. A. 2001, MNRAS, 321, 155
\bibitem[{Lokas et al.}(2011)]{lo11} {\L}okas, E. L., Kazantzidis, S., \& Mayer, L. 2011, ApJ, 739, 46
\bibitem[{Lokas et al.}(2012)]{lo12} {\L}okas, E. L., Majewski, S. R., Kazantzidis, S., et al. 2012,
	ApJ, 751, 61
\bibitem[{Lokas et al.}(2013)]{lo13} {\L}okas, E. L., Gajda, G., \& Kazantzidis, S. 2013, MNRAS, 433, 878
\bibitem[{Lokas \& Semczuk}(2014)]{ls14} {\L}okas, E. L., \& Semczuk, M. 2014,
in Proc. XXXVI Polish Astronomical Society Meeting, ed. A. R\'{o}\.{z}a\'{n}ska, \& M. Bejger
(Warsaw, Poland: PAS), 53
\bibitem[{Lokas et al.}(2014)]{lo14a} {\L}okas, E. L., Athanassoula, E., Debattista, V. P., et al. 2014a,
	MNRAS, 445, 1339
\bibitem[{Lokas et al.}(2014)]{lo14b} {\L}okas, E. L., Ebrov\'{a}, I., del Pino, A., \& Semczuk, M. 2014b,
	MNRAS, 445, L6
\bibitem[{Mayer et al.}(2001)]{may01} Mayer, L., Governato F., Colpi, M., et al. 2001, ApJ, 559, 754
\bibitem[{Navarro et al.}(1997)]{nfw97} Navarro, J. F., Frenk, C. S., \& White, S. D. M. 1997, ApJ, 490, 493 (NFW)
\bibitem[{Pawlowski \& Kroupa}(2014)]{pk14} Pawlowski, M. S., \& Kroupa, P. 2014, ApJ, 790, 74
\bibitem[{Read et al.}(2006)]{read06} Read, J. I., Wilkinson, M. I., Evans, N. W., Gilmore, G., \& Kleyna J. T.
	2006, MNRAS, 366, 429
\bibitem[{Springel}(2005)]{spr05} Springel, V. 2005, MNRAS, 364, 1105
\bibitem[{Springel et al.}(2001)]{syw01} Springel, V., Yoshida, N., \& White, S. D. M. 2001, New Astronomy, 6, 79
\bibitem[{Toomre \& Toomre}(1972)]{tt72} Toomre, A., \& Toomre, J. 1972, ApJ, 178, 623
\bibitem[{Widrow \& Dubinski}(2005)]{wd05} Widrow, L. M., \& Dubinski, J. 2005, ApJ, 631, 838
\bibitem[{Widrow et al.}(2008)]{widrow08} Widrow, L. M., Pym, B., \& Dubinski, J. 2008, ApJ, 679, 1239


\end{thebibliography}
\end{document}